# Res-Dense Net for 3D Covid Chest CT-scan classification

Quoc-Huy Trinh[1], Minh-Van Nguyen[1], and Thien-Phuc Nguyen-Dinh[1]

Ho Chi Minh University of Science

{20120013,20127094}@student.hcmus.edu.vn, ndtphuclqd1306@gmail.com

**Abstract.** One of the most contentious areas of research in Medical Image Preprocessing is 3D CT-scan. With the rapid spread of COVID-19, the function of CT-scan in properly and swiftly diagnosing the disease has become critical. It has a positive impact on infection prevention. There are many tasks to diagnose the illness through CT-scan images, include COVID-19. In this paper, we propose a method that using a Stacking Deep Neural Network to detect the Covid 19 through the series of 3D CT-scans images . In our method, we experiment with two backbones are DenseNet 121 and ResNet 101. This method achieves a competitive performance on some evaluation metrics.

## 1 Introduction

The SARS-COV-2 virus has spread over the world, with a major increase in 2019 and 2020. This global sickness cost nearly all governments in every country across the world a tremendous amount of money in 2020 and the first half of 2021citecovid stat. Thousands of individuals are infected with this disease every day, making it the most hazardous sickness on the planet. Almost all medical photos and records are maintained in a computerized database nowadays. Furthermore, the number of doctors available to diagnose these medical data is restricted, particularly in the case of Covid-19.

With the increasing number of patients and a scarcity of doctors, practically all photographs in the digital database may be used for pre-diagnosis, which helps to speed up diagnosis and improves the doctor's accuracy. This is why, in order to avoid infection, a quick and reliable detection approach is required. [13]. COVID-19 has a very severe on the respiratory system of the human. The virus is harbored most commonly with little or no symptoms, but can also lead to rapidly progressive and often fatal pneumonia. With the patients who have COVID-19, the virus can lead negative the patients situation [13].

There are a variety of diagnostic procedures available, including CT scans, chest X-rays, and PCR. They give us with a 3-D perspective of organ creation using CT-scan recording. Due to the lack of overlapping tissues, convenient disease evaluation, and its location, CT scans also provide a more complete overview of the internal structure of the lung parenchyma. [4]. As an aspect, it provides a window into pathophysiology that could shed light on several stages of disease detection and evolution.

Radiologists report COVID-19 patterns of infection with typical features including ground glass opacities in the lung periphery, rounded opacities, enlarged intra-infiltrate vessels, and later more consolidations that are a sign of progressing critical illness [2].

Chest radiography's medical imaging features are typically utilized to detect abnormalities and disorders in tissue such as the brain or lungs. As a result, chest x-ray images or CT-scan images are frequently utilized to detect abnormalities in human tissues.

Patients capture multiple slices during CT-scan recording, but the number of doctors available to diagnose these data is insufficient. These are the reasons why Machine Learning and Deep Learning models must be used in order to facilitate and shorten the diagnostic process. This is why the purpose of the 3D-CT scans pictures classification challenge is to assess several strategies for reliably and efficiently classifying 3D-CT scans images.[2].

In this paper, we propose a method that uses Fine-tuning and ensemble Deep Neural Network backbones to classify 3D CT-scan images. With the combination of Deep Neural Network, the model can have higher performance on feature extraction and classification task. In this experiment, we use two backbones, DenseNet 121 and ResNet 101, for evaluating our method on the test dataset with 4355 samples.

We also introduce the Res-Dense net architecture, the experiment, and the assessment mechanism in this section of our proposal. Our strategy earns a competitive score on the training and testing method in this experiment. Furthermore, we suggest various improvements to our methods' performance, and the method can be used to tackle other difficulties in medical imaging, notably in CT-scan images.

## 2    Related Work

### 2.1    CT-scan images

The difference between the CT-Scan images and other medical images, CT-Scan images are created by a series of X-ray images, which are forms of radiation on the electromagnetic spectrum. In addition, as compared to X-ray images, CT scans can provide information on multiple angles of the tissue and show the status of the tissue in each frame of image sequences. Furthermore, the information of the tissue can be clearly depicted through a sequence of photographs taken at the same moment. [24].

## 2.2    Image Classification

Image classification is a task that attempts to classify the image by a specific label. In recent years, the development of computing resources leads to a variety of methods in Image classification. Many deep architectures have been proposed and get competitive results. Moreover, from the classification task, the researcher can use to localize the abnormal on the images, which is very important in the medical diagnosis [22].

## 2.3    CT-Scan COVID-19 image classification

The use of imaging data is illustrated to be a helpful method to diagnose Covid19. However, computed tomography (CT-Scan) gets a variety of signs and creates difficulty for the doctors. However, with the development of computing and computer vision, there are several methods are proposed to deal with this problem. The flourish of the Deep Learning and Transfer Learning methods are creating a beneficial impact on image classification. By using CNN, Deep Neural Network architectures to classify the Covid-19 CT-scan images, the accuracy of diagnosing achieves competitively. Nowadays, there are many challenges in CT-Scan classification to find the competitive approach and method to apply in a fast and accurate diagnosis to prevent the societal infection [9].

# 3    Dataset

The dataset we use is from MIA-COVID 19 dataset, which contains the Covid 3D-CT Scan images series from patients that have COVID 19 and patients that do not have COVID 19[3]. The dataset is split into folders. Each of them is a series of images when doing CT-Scan. All of the images are collected from COVID19-CT-Database. The dataset include the input sequence is a 3-D signal, consisting of a series of chest CT slices, i.e., 2-D images, the number of which is varying, depending on the context of CT scanning. The context is defined in terms of various requirements, such as the accuracy asked by the doctor who ordered the scan, the characteristics of the CT scanner that is used, or the specific subject's features, e.g., weight and age [1].

The COVID19-CT-Database (COV19-CT-DB) consists of chest CT scans that are annotated for the existence of COVID-19. Data collection was conducted in the period from September 1, 2020, to March 31, 2021. Data were aggregated from many hospitals, containing anonymized human lung CT scans with signs of COVID-19 and without signs of COVID-19 [1].

The COV19-CT-DB database consist of about 5000 chest CT scan series, which correspond to a high number of patients (> 1000) and subjects (> 2000). Annotation of each CT slice has been performed by 4 very experienced (each with over 20 years of experience) medical experts; two radiologists and two pulmonologists. Labels provided by the 4 experts showed a high degree of agreement(around

98%)[2].

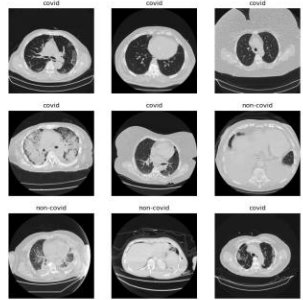

**Fig.1.** Sample of Dataset

### 3.1   Evaluation methods

To evaluate a classification model, we have some basic methods such as Accuracy, Precision, Recall, F1-score, etc. Regarding Precision, this measurement score evaluates the numbers of true positive over the number of false-positive and truepositive while Recall measure the number of positive over the true positive and false negative. However, with the F1-score, which is described as the harmonic mean of the two, the evaluation is similar to the average of Precision and Recall; the measurement score gets sensitive to two inputs having a low value, which helps to make the experiment fair [26].

$$\frac{Precision \times Recall}{Precision + Recall} \tag{1}$$

To evaluate the methods, we use the Macro F1-Score with the following formula[7]:

$$\frac{1}{n} * \sum_{i=0}^{n} F1 - scores_i \tag{2}$$

Where:

n: number of classes/labels i
: class/label

## 4    Method

Despite using CNN with RNN or LSTM to find features of the 3D CT-Scan series, we propose a method that extracts all features of all images in all series. This method can help efficiently reduce the time of training and achieve competitive performance. In the testing phase, we propose to predict all images in the series and calculate the mean score of the series to choose the label of that series.

### 4.1    Densely Connected Convolutional Network

The demonstration of the recent work has shown that Convolutional Neural Networks can be substantially deeper, more accurate, and more efficient to train if they contain shorter connections between each layer close to the input and those close to the output[20]. DenseNet connects each layer to every other layer in a feed-forward chain. Whereas traditional Convolutional Neural Network architecture with L layers has L connections - one between each layer and its subsequent layer - our network has $\frac{L(L+1)}{2}$ direct connections. For each layer, the feature maps of all preceding layers are used as inputs, and their feature maps are used as inputs into all subsequent layers. DenseNet have several compelling advantages: they alleviate the vanishing gradient problem, strengthens feature propagation, encourages feature reuse, and substantially reduces the number of parameters
[6].

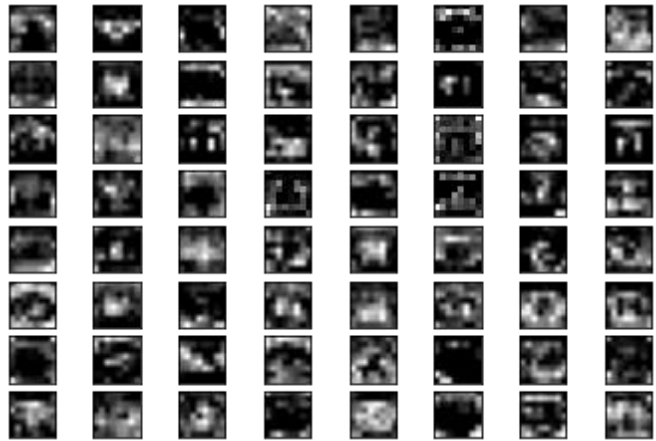

**Fig.2.** Feature Extraction of DenseNet 121

## 4.2    Deep Residual Network (ResNet)

Deeper neural networks are difficult to train. Therefore, Deep Residual Network is created to ease the training of networks that are substantially deeper than those used previously [11]. ResNet architecture can explicitly reformulate the layers as learning residual functions regarding the layer inputs, instead of learning unreferenced functions. With this network, it is easier to optimize and can gain accuracy from considerably increased depth. On the ImageNet dataset, the evaluation of residual nets, with a depth of up to 152 layers and eight times deeper than the VGG net, still has lower complexity. An ensemble of these residual nets achieves 3.57 errors on the ImageNet test set. This result won the first place on the ILSVRC 2015 classification task. The architecture also gains a competitive performance on the other dataset such as Cifar, etc [5].

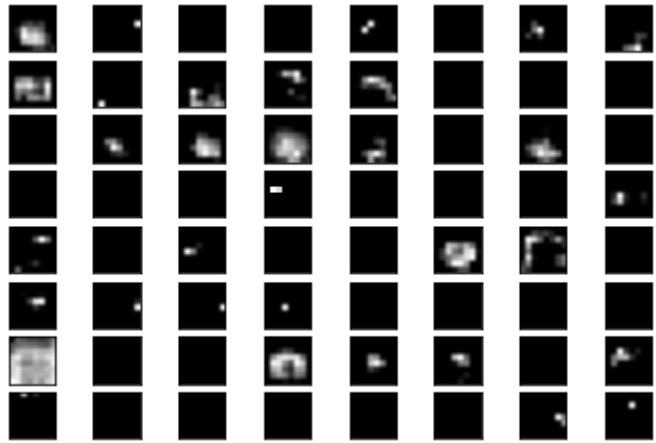

**Fig.3.** Feature Extraction of ResNet 101

## 4.3    Res-Dense Net

In our architecture, we propose to use ResNet 101 and DenseNet 121 backbones for the first layers by stacking techniques to create a new layer. This technique can use the merits of these two models to improve the strength and help reduce the drawback of the feature extraction process of two backbones.

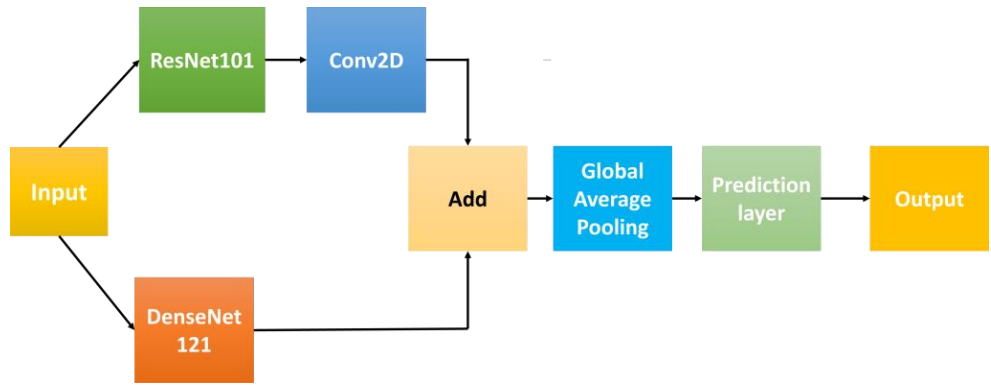

**Fig.4.** General Network Architecture

The input pass in two ways first is the Resnet 101 and a Convolution 2D layer, second, go through the DenseNet 121, then the result is added to the Add layer, then the result feature map then moves to the Global Average Pooling layer and the vector result of this layer is fed to the prediction layer for the output of the model. The figure below will illustrate obviously our work and the model that we design:

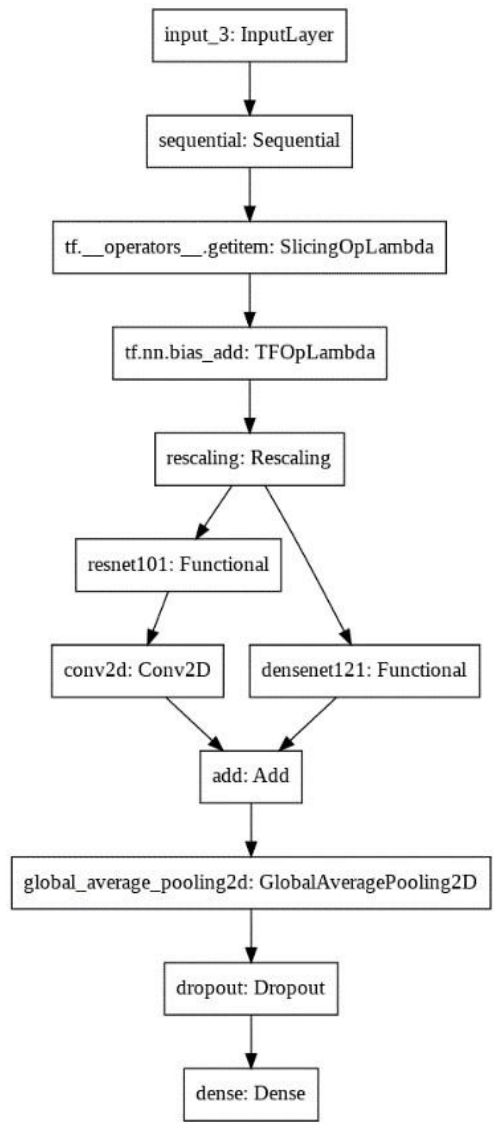

**Fig.5.** Full Network Architecture Visualization

One of the merits of our method is when feature maps, which are results of two previous extraction layers, are merged by adding layers to create the regular feature map and coming to the Global Average Pooling layers before coming to the Fully-Connected layer. By this technique, the importance of each area in the images can be defined by projecting weights of the output layer on the Convolution feature map that gain from the previous layer. Two Convolution Layers are added to make the output shape, and the number of filters of the two outputs from two

Deep Neural networks is equal. Moreover, with these layers, we can control the quality of output features that are created by the architecture.

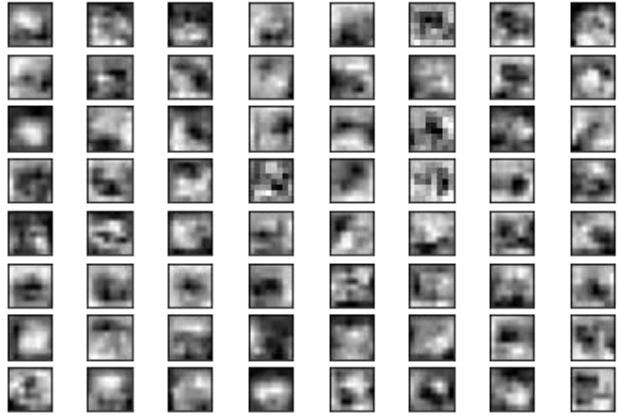

**Fig.6.** Feature extraction of Res-Dense Net

After that, features pass the Global Average Pooling layers and generate one feature map for each corresponding category of the classification task in the last MLP layer. With two backbones, new feature maps are created by two or more backbones. We propose to use global average pooling to reduce the computation cost.

The following is the Generalized Mean Pooling formula:

$$f^{(g)} = [f_1^{(g)} ... f_k^{(g)} ... f_K^{(g)}]^T, f_K^{(g)} = (\frac{1}{|\chi_k|} \sum_{x \in \chi_k} x^{P_k})^{\frac{1}{P_k}}$$

(3)

In the formula (3), the $f_{(g)}$ is the output of Convolution layers, each feature map from the output is calculated by the average and the result for each will be an element in the vector $f_i^{(g)}$ with $i = [1,k]$. From this formula, the number of nodes for feeding to the multi-layer perceptron reduces.

Instead of adding Flatten layers to create vectors for fully connected layers on top of the feature maps, this layer takes the average of each feature map. Then, the resulting vector is fed directly into the sigmoid activation function. The advantage of Global Average Pooling is there is no parameter to optimize in the global average pooling thus overfitting is avoided at this layer. Furthermore, global average pooling calculates the average out the spatial information, thus it is more robust to spatial translations of the input.

This architecture is inspired by the inception block of GoogleNet, by merging the convolutional layers and using Average pooling layers. This method helps the architecture get deeper but efficiently the computation cost.

### 4.4    Data preprocessing

After loading data, we resize all the images to the size (256,256), then we split the dataset into the training set and validation set in the ratio of 0.75:0.25. After resizing and splitting the validation set, we rescale the data pixel down to be in the range [-1,1]. Then we use the application of ResNet to preprocess the input. The Input after preprocess is rescaled to the same input of the ResNet model.

### 4.5    Data Augmentation

Data Augmentation is vital in the data preparation process. Data Augmentation improves the number of data by adding slightly modified copies of already existing data or newly created synthetic data from existing data to decrease the probability of the Overfitting problem, we use augmentation to generate the data randomly by random flip images and random rotation with an index of 0.2.

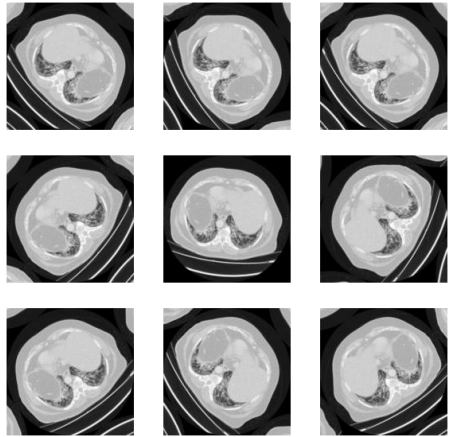

**Fig.7.** Data Augmentation result

### 4.6    Training

Our models are initialized with pre-trained weight from Imagenet. We use a batch size of 32 for training data with an image's size of (256,256). Moreover, we propose to use the RMSprop optimizer with a learning rate is 0.0001 for optimizer and evaluate the training process by accuracy and F1-score. For the loss function, we use Sparse Categorical Cross-entropy. The model is trained with 20 epochs and get the checkpoint that achieves the highest performance.

| Parameter | Value |
|---|---|
| Optimizer | Adam |
| Learning rate | 0.0001 |
| Loss | Binary Crossentropy |
| metrics | F1-score, accuracy |

**Table 1.** The parameter setup for model before training

With a low learning rate, we can ensure that we can find the weight with the competitive results although it costs more time for training.

Firstly, we freeze all the complicated layers of DenseNet and ResNet. Then, we start to train for the first time. Next, we freeze 100 layers before on each backbone. Finally, we start the continuous training process.

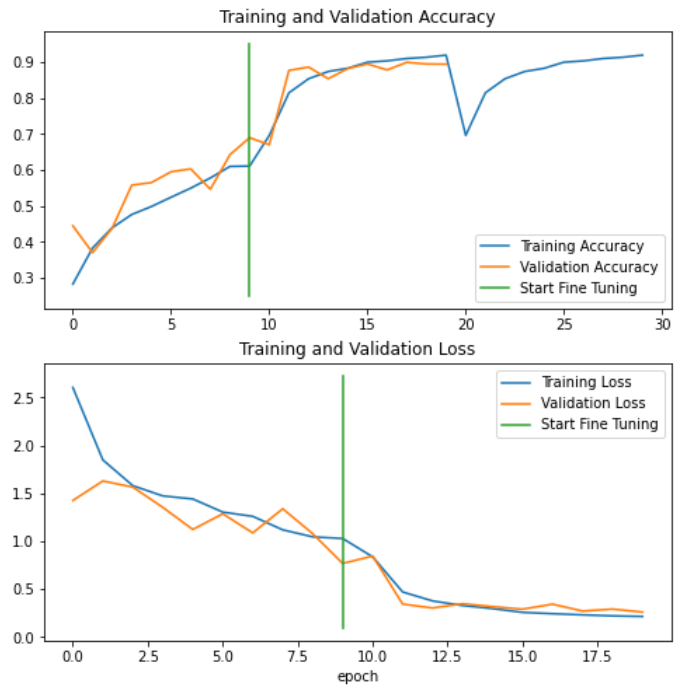

**Fig.8.** Training evaluation

After Fine-tuning with 20 epochs, we get the result of a loss of approximately 0.4. The model performs positively on the training dataset. On the evaluation of the Macro F1-score on the test set with 4355 folders of images that contain covid and non-covid, we achieve the competitive score:

| Score | Value |
|---|---|
| F1 (COVID) | 63.08 |
| F1 (NON-COVID) | 93.18 |
| Macro F1 | 78.13 |

**Table 2.** Our result on test set

## 5    Conclusion

We demonstrated the proposal of using Res-Dense Net with Fine-tuning technique to classify endoscopic images. The result of our research is positive for the F1 score. Moreover, our method can inspire a new approach on classification on 3D images instead of using 3D CNN or Convolution Neural Network with LSTM or RNN. However, there are some drawbacks that we have to do to improve the performance of the model, such as pre-processing data, reduce noise, change the size of the image to train. Furthermore, we can apply ResNet101 V2 or DenseNet 169 backbone, or we can combine with LSTM or RNN modules to have better feature extraction and better performance of the model.

## 6    Discussion

Our result get 78.13 in Macro F1-score, which gets a higher score than the Traditional 3D CNN and 3D Regnet [17] architecture [16]. However, our model gets the lower result with state of the art architecture[18]. Moreover, because of not increasing the quantity of data, our method also does not get a higher score, we can improve the accuracy of the model by improving data. The architecture that we combine is between ResNet 101 and DenseNet 121. Below is the result comparison between other methods with our method.

| Method | Macro F1-Score |
|---|---|
| 3D-CNN with Bert [18] | 88.22 |
| **Res-Dense Net** | **78.13** |
| 3D RegNet [16] | 71.83s |
| Shallow Convolution Neural-Network [19] | 70.86 |
| Vision Transformer [15] | 70.5 |

**Table 3.** Comparison with other models

With the approach of detecting Covid symptoms in each frame of the video, we get a higher result than others with the same approach but with the approach of using the sequence of frames in the video because the architecture uses the power of feature extraction from two backbones are ResNet 101 and DenseNet 121 create the benefit for the prediction process. Nevertheless, our method gets worse in this domain if compare with other methods using Bert or 3D CNN because this method can not find the relation of the frame like approaching sequences like 3D-CNN with Bert method or 3D RegNet method. However, our model has a better score than the 3D RegNet method that uses sequences of frames in the video.

## 7    Future Work

Although our method gets a competitive score, there are some drawbacks in our methods: the training time gets long with 334ms/step, we can custom layers in the architecture to accelerate the computing cost. We can get more layers or can ensemble more backbones to achieve higher results. Moreover, we can do a segmentation process on the lung CT-scan to improve the accuracy of training and get a better result for our architecture that is proposed [25].